\documentclass[12pt]{article}
\usepackage{amsfonts,amssymb}
\usepackage{graphics,psboxit,amsmath} %(if uncomment \be does not get a name)
\usepackage{subfigure}
\usepackage{graphicx}
\usepackage{verbatim}
\usepackage{young}
\def\hybrid{\topmargin -30pt    \oddsidemargin 0pt %%%%%%%%%%%%%% Archive-30pt
        \headheight 0pt \headsep 0pt
        \textwidth 6.25in       % A4 paper
        \textheight 9.5in       % A4 paper
        \marginparwidth .875in
        \parskip 5pt plus 1pt   \jot = 1.5ex}

\hybrid

\def\baselinestretch{1.2}

\catcode`\@=11

\def\marginnote#1{}
\newcount\hour
\newcount\minute
\newtoks\amorpm
\hour=\time\divide\hour by60
\minute=\time{\multiply\hour by60 \global\advance\minute by-\hour}
\edef\standardtime{{\ifnum\hour<12 \global\amorpm={am}%
        \else\global\amorpm={pm}\advance\hour by-12 \fi
        \ifnum\hour=0 \hour=12 \fi
        \number\hour:\ifnum\minute<10 0\fi\number\minute\the\amorpm}}
\edef\militarytime{\number\hour:\ifnum\minute<10 0\fi\number\minute}
\def\draftlabel#1{{\@bsphack\if@filesw {\let\thepage\relax
   \xdef\@gtempa{\write\@auxout{\string
      \newlabel{#1}{{\@currentlabel}{\thepage}}}}}\@gtempa
   \if@nobreak \ifvmode\nobreak\fi\fi\fi\@esphack}
        \gdef\@eqnlabel{#1}}
\def\@eqnlabel{}
\def\@vacuum{}
\def\draftmarginnote#1{\marginpar{\raggedright\scriptsize\tt#1}}

\def\draft{\oddsidemargin -.5truein
        \def\@oddfoot{\sl preliminary draft \hfil
        \rm\thepage\hfil\sl\today\quad\militarytime}
        \let\@evenfoot\@oddfoot \overfullrule 3pt
        \let\label=\draftlabel
        \let\marginnote=\draftmarginnote
   \def\@eqnnum{(\theequation)\rlap{\kern\marginparsep\tt\@eqnlabel}%
\global\let\@eqnlabel\@vacuum}  }

\def\draft2{
        \def\@oddfoot{\sl preliminary draft \hfil
        \rm\thepage\hfil\sl\today\quad\militarytime}
        \let\@evenfoot\@oddfoot \overfullrule 3pt
        \let\label=\draftlabel
        \let\marginnote=\draftmarginnote
   \def\@eqnnum{(\theequation)\rlap{\kern\marginparsep\tt\@eqnlabel}%
\global\let\@eqnlabel\@vacuum}  }

\def\preprint{\twocolumn\sloppy\flushbottom\parindent 2em
        \leftmargini 2em\leftmarginv .5em\leftmarginvi .5em
        \oddsidemargin -.5in    \evensidemargin -.5in
        \columnsep .4in \footheight 0pt
        \textwidth 10.in        \topmargin  -.4in
        \headheight 12pt \topskip .4in
        \textheight 6.9in \footskip 0pt
        \def\@oddhead{\thepage\hfil\addtocounter{page}{1}\thepage}
        \let\@evenhead\@oddhead \def\@oddfoot{} \def\@evenfoot{} }

\def\numberbysection{\@addtoreset{equation}{section}
        \def\theequation{\thesection.\arabic{equation}}}

\def\underline#1{\relax\ifmmode\@@underline#1\else
        $\@@underline{\hbox{#1}}$\relax\fi}

\def\titlepage{\@restonecolfalse\if@twocolumn\@restonecoltrue\onecolumn
     \else \newpage \fi \thispagestyle{empty}\c@page\z@
        \def\thefootnote{\fnsymbol{footnote}} }

\def\endtitlepage{\if@restonecol\twocolumn \else \newpage \fi
        \def\thefootnote{\arabic{footnote}}
        \setcounter{footnote}{0}}  %\c@footnote\z@ }

\catcode`@=12
\relax

\newcommand{\be}{\begin{equation}}
\newcommand{\ee}{\end{equation}}
\def\bea{\begin{eqnarray}}
\def\eea{\end{eqnarray}}

\def\nn{\nonumber}
\def\del{\partial}
\def\no{\noindent}

\def\gh{{\hat\gamma}}

\newcommand{\beq}{\begin{equation}}
\newcommand{\eeq}{\end{equation}}
\newcommand{\ben}{\begin{eqnarray}}
\newcommand{\een}{\end{eqnarray}}
\newcommand{\bes}{\begin{subequations}}
\newcommand{\ees}{\end{subequations}}
\newcommand{\blg}{\begin{align}}
\newcommand{\elg}{\end{align}}
\newcommand\diff{\mathrm{d}}
\newcommand{\half}{{\frac{1}{2}}}

\newcommand{\cN}{{\cal N}}
\newcommand{\startappendix}{
\setcounter{section}{0}
\renewcommand{\thesection}{\Alph{section}}}

\newcommand{\Appendix}[1]{
\refstepcounter{section}
\begin{flushleft}
{\large\bf Appendix \thesection: #1}
\end{flushleft}}

\def\N{{\cal N}}

\def\qq{\qquad}
\def\one{\mbox{1 \kern-.59em {\rm l}}}

\def\a{\alpha}      
       
  \def\G{\Gamma}  
\def\d{\delta}  \def\D{\Delta}

\def\l{\lambda} \def\L{\Lambda}

\def\r{\rho}
\def\s{\sigma}  

\def\th{\theta} \def\Th{\Theta}

  \def\cI{{\cal I}}
\def\cJ{{\cal J}}  
 \def\cN{{\cal N}} \def\cO{{\cal O}}

 \def\cZ{{\cal Z}}

\def\const{{\rm const.}}
\begin{document}

\renewcommand{\theequation}{\thesection.\arabic{equation}}
\csname @addtoreset\endcsname{equation}{section}

\newcommand{\eqn}[1]{(\ref{#1})}
%\draft2

\begin{flushright}
April 2012 \hfill
WITS-CTP-090\\
FPAUO-12/03\\
CPHT-RR010.0212
\end{flushright}

\vspace{16pt}

%%%%%%%%%%%%%%%%%%%%%%%%%%%%

\centerline{\Large \bf
Non-singlet Baryons in Less Supersymmetric Backgrounds}
\vspace{0.7cm}

\centerline{
{\large  Dimitrios Giataganas${}^{a,}$}\footnote{
                                   {\tt dimitrios.giataganas$@$wits.ac.za}},
\ {\large  Yolanda Lozano${}^{b,}$}\footnote{
                                   {\tt ylozano$@$uniovi.es } }}

\vspace*{0.3cm}
\centerline{
 {\large Marco Picos${}^{b,}$}\footnote{{\tt picosmarcos$@$uniovi.es}},
\  {\large  Konstadinos Siampos${}^{c,}$}\footnote{
                                   {\tt ksiampos$@$cpht.polytechnique.fr.}}
 }

\vspace{.3cm}

\centerline{{\it ${}^a$ National Institute for Theoretical Physics,
School of Physics }}
\centerline{{\it and Centre for Theoretical Physics,  }}
\centerline{{\it University of the Witwatersrand, Wits, 2050, South Africa}}

\vspace{.4cm}
\centerline{{\it ${}^b$Department of Physics,  University of Oviedo,}}
\centerline{{\it Avda.~Calvo Sotelo 18, 33007 Oviedo, Spain}}

\vspace{.4cm}
\centerline{{\it ${}^c$ Centre de Physique Th\'eorique, Ecole Polytechnique, CNRS }}
\centerline{{\it 91128 Palaiseau, France}}

\vspace{30pt}

%%%%%%%%%%%%%%%%%
\centerline{\bf ABSTRACT}
\vspace{.5truecm}

\noindent
We analyze the holographic description of non-singlet baryons in various backgrounds with reduced supersymmetries and/or confinement. 
We show that they exist in all $AdS_5\times Y_5$ backgrounds with $Y_5$ an Einstein manifold bearing five form flux, for a number of 
quarks $5N/8< k\leqslant N $,  independently on the supersymmetries preserved. This result still holds for $\gamma_i$ deformations.
 In the confining Maldacena-Nu\~nez background non-singlet baryons also exist, although in this case the interval for the number 
of quarks is reduced as compared to the conformal case.  
We generalize these configurations to include a non-vanishing magnetic flux such that a complementary microscopical
description can be given in terms of lower dimensional branes expanding into fuzzy baryons. This description is a first step towards exploring
the finite 't Hooft coupling region.\\

%We analyze the holographic description of baryons in ${\cal N}=1$ supersymmetric 
%gauge theories. We consider $Y^{p,q}$ dual spaces,
%$\beta$ (and $\gamma_i$) deformed backgrounds  and the Maldacena-Nu\~nez model,
%such that we have a sample of less supersymmetric backgrounds with different conformal behaviors. We find that for all conformal backgrounds
%there exist classical solutions representing bound states of $k$ quarks with $k\leqslant N$, i.e. non-singlet baryons, which are stable beyond the
%same critical value of $k$ found for ${\cal N}=4$ SYM. In the Maldacena-Nu\~nez background these states also exist, but for a more reduced interval
%for the number of quarks. This result points at a universal behavior of holographic non-singlet baryons based on conformality.
%We generalize these configurations to include an instantonic magnetic flux such that a complementary microscopical
%description can be given in terms of lower dimensional branes expanding into fuzzy baryons. This description is a first step towards exploring
%the finite 't Hooft coupling region.\\

\setcounter{page}0
\setcounter{footnote}0
\newpage

\tableofcontents

\def\baselinestretch{1.2}
\baselineskip 20 pt

\section{Introduction}

Baryon configurations were first suggested in the context of the AdS/CFT \cite{maldacena98} correspondence in \cite{witten98,GO}. The
gravitational dual of a bound state of  $N$ static external quarks in ${\cal N}=4$ SYM, the so-called baryon vertex, was found in terms of a
D5-brane wrapping the $S^5$ part of the spacetime geometry \cite{witten98}. If the $D5$-brane is point-like in the $AdS_5$ space, its Chern--Simons (CS)
action is a tadpole term which can be canceled if we introduce Chan--Paton factors for N-strings, whose endpoints at the boundary of $AdS$ represent the $N$ external quarks.
The classical solution corresponding to
this configuration was found in \cite{Brandhuber:1998,Imamura1} using
a generalization of the techniques in \cite{Malda} for the heavy quark-antiquark system. In this approach the influence of the
F-strings has to be considered in
order to analyze the stability of the baryon vertex in the holographic AdS direction. The energy of the system is then inversely proportional
to the distance between the quarks and since the proportionality constant is negative the configuration is stable in the AdS direction.

The description in \cite{Brandhuber:1998,Imamura1} suffices to deduce the basic properties of the system. However strictly speaking it is only valid when the endpoints of the $N$ F-strings are uniformly
distributed on the $S^5$, so that the latter is not deformed and the probe brane approximation holds. In this
approximation all supersymmetries are broken, and this results in a non-vanishing binding energy.  In order
to have some supersymmetries preserved all strings should end on a point, and then the deformation caused by
their tensions and charges should be taken into account. Incorporating the gauge field on the brane the
binding energy becomes zero, reflecting the fact that the configuration is supersymmetric \cite{Imamura2}.

The usual baryon refers to a bound state of N-quarks which form the completely antisymmetric representation of $SU(N)$.
In the holographic description however it is possible to construct a bound state of $k$-quarks with $k<N$ (Figure 1).
The bound state consists of a D5 or D3-brane wrapping the internal space\footnote{A submanifold of it in the case of the D3-brane.} located in the bulk, $k$ strings stretched between the brane and the
boundary of $AdS$  representing the quarks, and $N-k$ straight strings that go from the D5 or D3 brane deeper in the
bulk to a minimum distance. The bound on how low can the $k$ number go  depends on a no-force condition along the AdS
direction, and a priori seems to be affected by the geometry of both the internal and the
$AdS$ spaces. In the $AdS_5 \times S^5$ background $k$ should satisfy $5N/8 < k \leqslant N$ \cite{Brandhuber:1998,Imamura1}.
A stability analysis against fluctuations shows that the configurations are stable for a more restricted number of quarks
$0.813 N \leqslant k \leqslant N$ \cite{SS2}. An interesting question is what happens to the bound when the supersymmetry is reduced or
the conformal invariance is broken and more particularly if confinement is present. A physical expectation would be that at least the lower bound should
increase. One of the motivations of this paper is to investigate the 
bound dependence on the supersymmetry and confinement properties of the gauge theory.

Baryon vertex configurations in $AdS_5 \times T^{1,1}$ \cite{t11} and $AdS_5\times Y^{p,q}$ \cite{se,gauntlett06,benvenuti04}
geometries have been considered
in \cite{ACR} and \cite{CEPRV}, respectively. Using the full DBI description it has been shown that they are non-supersymmetric. General
properties of baryons in the Klebanov-Strassler \cite{KS} and Maldacena-Nu\~nez \cite{MN} models have also been discussed in \cite{Loewy:2001pq} (see also \cite{HP, Imamura3}). In these confining backgrounds the baryon is also non-supersymmetric and
is significantly different than in the previous cases, with an energy  linearly proportional to its size.

In this paper we analyze the dynamics of non-singlet baryons  in some of these backgrounds in the probe brane approach. 
We show that stable configurations exist with non-zero binding energy
as long as the number of quarks $k$ satisfies $k_{\rm min} < k \leqslant N$. The value of $k_{\rm min}=5N/8$ for all $AdS_5 \times Y_5$ backgrounds with $Y_5$ an Einstein manifold bearing five-form flux, and also for multi-$\beta$ deformed spaces \cite{LM, Frolov}. The analysis on the
deformed spaces basically gives the same undeformed results of $\N=4$ SYM. This is not unexpected since classical properties like energy and temperature, string
configurations, like the 1/4 BPS like Wilson loop, and brane configurations like particular giant gravitons
remain also non trivially undeformed \cite{Avramis:2007wb, giataganaswl, Im}. 
A stability analysis  confirms that the configurations
are stable for a number of quarks  $0.813 N \leqslant k \leqslant N$, again the same interval found  for the $AdS_5\times S^5$
background \cite{SS2}. These findings seem to contradict our expectations that non-singlet states should be more constrained in
theories with reduced supersymmetry. Rather, their existence seems to be
quite universal and independent on the amount of supersymmetries preserved. We should however keep in mind that the approach taken here breaks all the supersymmetries (see the conclusions for a further discussion on this point).
The same analysis for the $\cN=1$ Maldacena-Nu\~nez background \cite{MN} confirms that
non-singlet holographic baryons also exist in confining theories. However
broken conformal invariance and more particularly confinement increases the minimum number of quarks.

More general baryon vertex configurations with a non-vanishing magnetic flux have been suggested as a first step towards
accessing the finite 't Hooft coupling region in the dual CFT \cite{Janssen,GLR}. Indeed, showing that these configurations exist for finite $\lambda$ is of special interest when they are not BPS. 
Allowing for a non-trivial magnetic
flux has the effect of adding lower dimensional brane charges to the configuration. This in turn hints at the existence
of a microscopical description in terms of non-Abelian lower dimensional branes expanding into the baryon vertex by
means of Myers dielectric effect \cite{Myers}. This description allows to explore the configuration in the region
$R<<n^{1/(r-p)}\, l_s$, where $p$ is the dimensionality and $n$ the number of  expanding branes and $r$ the dimensionality of the resulting expanded brane, and is therefore
complementary to the supergravity description in terms of probe branes. Thus it is a first step towards exploring
the finite 't Hooft coupling region of the dual CFT from the gravity side.

The paper is organized as follows. We start in section 2 with a brief review of the holographic description of baryon vertices and their stability under small fluctuations for a general class of backgrounds. In section 3 we use these results to study the dynamics of the baryon vertex
in $AdS_5\times Y_5$, with $Y_5$ an Einstein manifold bearing five-form flux. We particularize to
the $AdS_5 \times Y^{p,q}$ and $AdS_5\times T^{1,1}$ geometries, where we switch on a non-vanishing magnetic flux suitable for the microscopical description of the $T^{1,1}$ in section 6.
 In section 4 the multi-$\beta$-deformed Frolov's background is considered. 
In section 5 we analyze the Maldacena-Nu\~nez 
background, where we confirm the existence of non-singlet baryons for a more constrained interval for $k$ due to confinement. We show that
in this case the stability requirement does not reduce the allowed interval.
In section 6 we perform the microscopical analysis,
in terms of D1 or D3-branes, depending on the background.
We identify the CS
couplings responsible for the F-string tadpoles of the configurations. In section 7 we summarize our results and discuss further directions. Finally, in the appendix we
collect some properties of the $Y^{p,q}$ and $T^{1,1}$ geometries relevant for our analysis and address the microscopical description of the baryon vertex in the $Y^{p,q}$ geometries. 

\begin{figure}[!t]
\begin{center}
\begin{tabular}{cc}
\includegraphics[height=10cm]{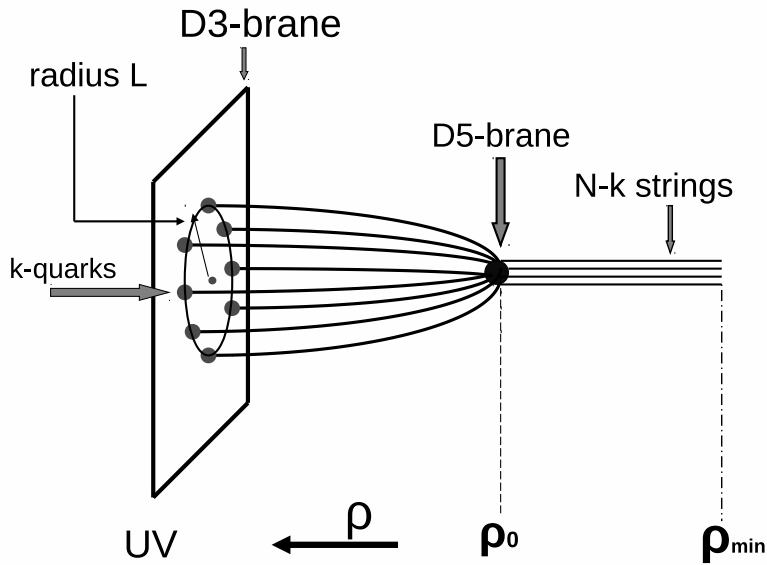}
\end{tabular}
\end{center}
\vskip -2.5 cm \caption{A baryon configuration with $k$-external quarks placed on a
spherical shell of radius $L$
at the boundary of AdS space, each connected to the wrapped $Dp$-brane located at $\rho=\rho_0$ and $N-k$ straight
strings ending at $\rho_{\rm min}$.}
\end{figure}

\section{The holographic baryon vertex construction}

In this section we review the holographic description of baryons in the general class of backgrounds presented in \cite{SS2}, as well as the study of their stability against small fluctuations. The first part generalizes the construction in \cite{Brandhuber:1998,Imamura1} to non-conformal cases like the Maldacena-Nu\~nez background that we will discuss in section 5.

We consider diagonal metrics of Lorentzian signature of the form
\bea
\label{metric}
ds^2 = G_{tt} dt^2 + G_{xx}(dx^2 + dy^2 + dz^2)+G_{\rho\rho}d\rho^2 + R^2d\mathbb{M}_p^2 \ , 
\eea
where $x,y$ and $z$ denote cyclic coordinates and $\rho$ denotes the radial direction playing the
role of an energy scale in the dual gauge theory. It extends
from the UV at $\rho \to \infty$ down to the IR at some minimum value
$\rho_{\rm min}$ determined by the geometry.

\no
It is convenient to introduce the functions
\begin{equation}
%\label{2-2}
f(\rho) = - G_{tt}G_{xx}\ ,\qquad g(\rho)=- G_{tt} G_{\rho\rho}\ , \qquad h(\rho)=G_{xx}G_{\rho\rho} \ ,
\end{equation}
which for $AdS_5\times\mathbb{M}_5 $ with radii $R$  read
\begin{equation}
%\label{2-3}
 f(\rho)= \rho^4\ ,\qquad g(\rho)= 1\ ,\qquad h(\rho)= 1\ .
\end{equation}

As we have mentioned, a non-singlet baryon is described holographically in terms of a $Dp$-brane wrapping the internal manifold $\mathbb{M}_p$ with $k$ fundamental strings connecting it to the boundary at $\rho\rightarrow\infty$. The remaining $N-k$ straight strings go from the $Dp$-brane straight up at 
$\rho_{\rm min}$. The binding potential energy of the baryon is then given by
$e^{-{\rm i} E T} = e^{{\rm i} S_{\rm cl}}$,
where $S_{\rm cl}$ is the classical action of the holographic baryon.
\no
This action consists of three terms,
the Nambu--Goto action for the strings stretching from the baryon vertex to the boundary at $\rho\rightarrow\infty$, the Nambu--Goto action for the straight strings stretching between the brane and $\rho_{\rm min}$ and the 
Dirac--Born--Infeld action for the $Dp$-brane
\bea
%\label{2-5}
&&S_{F1} = - {1\over 2 \pi} \int d \tau d \sigma \sqrt{-\det P (G_{\alpha \beta})}\ ,
\nonumber\\
&& S_{Dp}^{DBI}=-T_p \int_{\mathbb{R}\times \mathbb{M}_p}  d^{p+1}\xi\, \sqrt{-\det P(G_{ab}+2 \pi F_{ab}-B_{ab})}\ ,
\nonumber
\eea
where $F$ is the Born-Infeld field strength.

\no
We first fix reparametrization invariance for each string by choosing
\bea
%\label{2-6}
t=\tau \ ,\qq \rho=\sigma \ .
\eea
For static solutions
we consider the embedding of the $S^2$--sphere on the $D3$--brane in spherical
coordinates $(r,\th,\phi)$
\bea
%\label{2-7}
r = r(\rho)\ ,\qquad (\th, \phi) =\const\ ,
\eea
plus $\mathbb{M}_p\hbox{--angles} = \const$, 
supplemented by the boundary condition
\bea
%\label{2-8}
\rho \left(L \right)=\infty\ .
\eea
Then, the Nambu--Goto action for the strings stretching from the baryon vertex to the boundary of AdS reads
\bea
\label{NG action}
S = - {T\over 2 \pi} \int_{\rho_0}^\infty d\rho\, \sqrt{ g(\rho) + f(\rho) r^{\prime 2}}\ ,
\eea
where $T$ denotes time and the prime denotes a
derivative with respect to $\rho$. From the Euler--Lagrange equations of motion
we obtain
\bea
\label{eom}
{f r_{\rm cl}^\prime \over \sqrt{g+ f r_{\rm cl}^{\prime 2}}}
=  f_1^{1/2}\qquad \Longrightarrow
\qquad r_{\rm cl}^\prime =  {\sqrt{f_1F}\over f}\ ,
\eea
\no
where $\rho_1$ is the value of $\rho$ at the turning point of each string,
$f_1 \equiv f(\rho_1),\ f_0\equiv f(\rho_0)$ and
\bea
%\label{2-11}
F = {g f \over f - f_1}\ .
\eea
The $N-k$ strings which extend from the baryon vertex to $\rho=\rho_{\rm min}$
are straight, since $r^\prime=0$ is a solution of the equations of
motion (with $f_1=0$) and satisfies the boundary condition at the vertex.
Integrating \eqn{eom} we can express the radius of the spherical shell as
\bea
\label{length}
L =\sqrt{f_1}\int_{\rho_0}^{\infty} d \rho {\sqrt{F}\over f}\ .
\eea
\no
Next we fix the reparametrization invariance for the wrapped $Dp$-brane by choosing
\bea
\label{DBI.gauge}
 t=\tau\ ,\qq \th_a=\s_\a\ , \quad \a=1,2,\dots,p\ .
\eea

Finally, inserting the solution for $r_{\rm cl}^\prime$ into \eqn{NG action} and
subtracting the divergent energy of its constituents we can write the binding energy of the baryon as
\begin{equation}
\label{energy}
E = {k \over 2\pi}\left\{\int_{\rho_0}^\infty d \rho \sqrt{F}
- \int_{\rho_{\rm min}}^\infty d \rho \sqrt{g}
+{1-a\over a} \int_{\rho_{\rm min}}^{\rho_0} d \rho \sqrt{g}\ +\frac{2\pi}{ a N}\,E_{Dp}\bigg|_{\rho=\rho_0} \right\}\ ,
\end{equation}
where
\bea
a\equiv {k\over N}\ ,\qq 0< a \leqslant 1\ .
\eea
The expressions for the length and the energy, \eqn{length} and \eqn{energy}, depend on the arbitrary
parameter $\rho_1$ which should be expressed in terms of the baryon vertex position $\rho_0$.
The most convenient way to find this is to impose that the net force at the baryon vertex is zero \cite{SS1,SS2}
\bea
\label{force}
&&\cos\Th = {1-a\over a} +\frac{2\pi}{a N}\frac{1}{\sqrt{g}}\, \partial_\rho E_{Dp}\bigg|_{\rho=\rho_0}\ ,
\\
&& \cos\Th =  \sqrt{1-f_1/f_0}\ ,\nonumber
\eea
where $\Th$ is the angle between each of the
$k$-strings and the $\rho$-axis at the baryon vertex,
which determines $\rho_1$ in terms of $\rho_0$.
An alternative derivation of this expression can be found by demanding that the physical 
length \eqn{length} does not depend on the arbitrary parameter $\r_1$, in other words
\bea
\label{derivative}
\frac{\del L}{\del \r_1}=0\Longrightarrow
\frac{\del \r_0}{\del \r_1}=\frac{f_1'}{2\tan\Th}\sqrt{\frac{f_0}{g_0f_1}}
\int_{\r_0}^\infty d\r\frac{\sqrt{gf}}{(f-f_1)^{3/2}}\ .
\eea
Minimizing the energy \eqn{energy} with respect to $\r_1$ and using \eqn{derivative}
we find the no-force condition \eqn{force}.
 Using (\ref{length}), (\ref{energy}) and (\ref{force}) it is also possible to see that
\begin{equation}
\label{Evar}
\frac{dE}{d\rho_0}=\frac{k\sqrt{f_1}}{2\pi}\frac{dL}{d\rho_0}
\end{equation} 
which will be useful when we study the Maldacena-Nu\~nez background.

As we will see in the examples to follow, \eqn{force} has a solution
for a parametric region of $(a,\rho_0)$. 
However, in order to isolate parametric regions
of physical interest a stability analysis of the classical solution should be performed, which further restricts the allowed region. We know from \cite{SS2} that instabilities can only emerge from longitudinal fluctuations of the $k$ strings, since only these possess a non-divergent zero mode, which is a sign of instability. To study the fluctuations about the classical solution the embedding should be perturbed according to
\begin{equation}
r=r_{cl}+\delta r(\rho)\, ,
\end{equation}
and the Nambu-Goto action should be expanded to quadratic order in the fluctuations. $\delta r$ is then solved from the equation
\begin{equation}
\label{stabeq1}
\frac{d}{d\rho}\Bigl(\frac{gf}{F^{3/2}}\frac{d}{d\rho}\Bigr)\delta r=0
\end{equation}
This has to be supplemented with the boundary condition for the $\delta r$ fluctuations, given by equation (3.12) in \cite{SS2}
\begin{equation}
\label{stabeq2}
2(f-f_1)\delta r^\prime+\delta r\, \Bigl(2f^\prime-\frac{f^\prime}{f}f_1-\frac{g^\prime}{g}(f-f_1)\Bigr)=0 \qquad {\rm at}\,\,\, \rho=\rho_0
\end{equation}
As we will see in the examples to follow these conditions further restrict the parametric region $(a,\rho_0)$ for which a classical non-singlet baryon solution exists.

%%%%%%%%%%%%%%%%%%%%%%%%%%%%%%%%%%%%%%%%%%%%%%%%%%%%%%%%%%%%%%%%%%%%%%%%%%%%%%%%%%%%%%%%%%%%%%%%%%%%%%%%%%%%

\section{The baryon vertex in $AdS_5\times Y_5$ manifolds}

The holographic description of the baryon vertex in  $AdS_5\times Y_5$ backgrounds with $Y_5$ an Einstein manifold bearing five-form flux is identical, in the probe brane approximation, to that in $AdS_5\times S^5$ \cite{Brandhuber:1998, Imamura1}. Therefore  
non-singlet
states exist for the same number of fundamental strings $5N/8< k\leqslant N$. Spike solutions associated to the baryon vertices in the $AdS_5 \times Y^{p,q}$ and  $AdS_5\times T^{1,1}$ geometries  have been discussed in  \cite{CEPRV} and \cite{ACR}, where it has been shown that they break all the supersymmetries. Therefore we are certain that the bound states found in the probe brane approximation will not become marginal due to supersymmetry once the backreaction is taken into account.
In these two geometries we will switch on a magnetic flux that will dissolve D1 and D3-brane charges in the configuration. The vertex will then be described at finite 't Hooft coupling in terms of D1-branes expanding into a fuzzy $S^2\times S^2$ submanifold of the $T^{1,1}$ for the Klebanov-Witten background and D3-branes expanding into a fuzzy $S^2$ submanifold of the $Y^{p,q}$ for the Sasaki-Einstein. The detailed microscopical analysis of these configurations will be performed in section 6 and the appendix respectively.

\subsection{The D5-brane baryon vertex}

In our conventions the $AdS_5\times Y_5$ metric reads
\begin{equation}
ds^2=\frac{\rho^2}{R^2}dx^2_{1,3}+\frac{R^2}{\rho^2}d\rho^2+R^2 ds^2_{Y_5}~,
\end{equation}
with $R$ the radius of curvature in string units,
\begin{equation}
\label{radiusYpq}
R^4=\frac{4\pi^4Ng_s}{{\rm Vol}(Y_5)}~.
\end{equation}
The $AdS_5\times Y_5$ flux is given by $F_5=(1+\star_{10}){\cal F}_5$, where
\begin{equation}
{\cal F}_5=4\, R^4\, d{\rm Vol}(Y_5)\ .
\end{equation}

A D5-brane wrapping the whole $Y_5$ captures the $F_5$ flux, and it requires the addition of $N$ fundamental strings to cancel the tadpole
\begin{equation}
\label{tadpole}
S^{CS}_{D5}=2\pi\, T_5 \int_{\mathbb{R}\times Y_5} P[C_4] \wedge F=
-2\pi\, T_5 \int_{\mathbb{R}\times Y_5} P[F_5] \wedge A=-N \int dt A_t~,
\end{equation}
where $A$ is the Born-Infeld vector field.
The DBI action is in turn given by
\begin{equation}
S^{DBI}_{D5}= -T_5 \int_{\mathbb{R}\times Y_5}  d^6\xi \, e^{-\phi} \sqrt{-{\rm det} P(G)}
=-\frac{TN }{8\pi}\rho_0\ .
\end{equation}

\subsubsection{Classical solution}

Given that the energy of the D5-brane is independent of the volume of the Einstein manifold the classical solution in the probe brane approximation is the one found in \cite{Brandhuber:1998, Imamura1} for $AdS_5\times S^5$. Making contact with the analysis in the previous section we now have
\begin{equation}
-G_{tt}=G_{xx}= G_{\rho\rho}^{-1}={\rho^2\over R^2}\ .
\end{equation}
The radius and the energy are then given in terms
of the
position of the $D5$-brane $\rho_0$ and the turning point $\rho_1$ of each string, as
\bea
L={R^2\rho_1^2\over 3\rho_0^3}\ {\cal I}\ ,\qq E={k \rho_0\over 2\pi}\left(- {\cal J}+{5-4a\over 4a }\right)\ ,
\eea
with ${\cal I}$, ${\cal J}$ the hypergeometric functions
\bea
{\cal I}= {_2F_1}\left({1\over 2},{3\over 4},{7\over 4};{r_1^4\over r_0^4}\right)\ ,\qq
{\cal J}={_2F_1}\left(-{1\over 4},{1\over 2},{3\over 4};{r_1^4\over r_0^4}\right)\, ,
\eea
exactly as in $AdS_5\times S^5$.
From \eqn{force} we find that the no-force condition on the $\rho$-axis yields
\bea
\label{turning}
\rho_1=\rho_0(1-\l^2)^{1/4}\ ,\qq \l={5-4a\over 4a}\ , \qq a\equiv{k\over N}\ .
\eea
Given that $\l<1 $ a baryon
configuration exists for
$a >  a_<$ with 
$a_<={5\over8}$ 
\noindent
Finally, the binding energy in terms of the physical length of the baryon reads
\bea
E=-{R^2\over2\pi L}{k\sqrt{1-\l^2}\over 3}\left({\cal J}-
{5-4a\over 4 a}\right){\cal I}\ .
\eea
Thus it has both the expected behavior with $1/L$ dictated by conformal invariance and the non-analyticity of square-root branch cut type in the 't Hooft parameter 
\cite{Malda,Brandhuber:1998,GLR}. We would also like to point out that our string and brane configurations satisfy the Sasaki-Einstein constrains in the way studied in \cite{giataganasSE} and therefore our solutions are in this sense valid.

\subsubsection{Stability analysis}

Again as in $AdS_5\times S^5$ \cite{SS2}
the study of the stability against longitudinal fluctuations gives
\bea
\label{zeromode}
\d r(\rho)= A \int_\rho^\infty d\rho{\rho^2\over(\rho^4-\rho_1^4)^{3/2}}={A\over 3\rho^3}\ {_2F_1}
\left({3\over 4},{3\over 2},{7\over 4};{\rho_1^4\over \rho^4}\right)
\eea
as the solution of equation (\ref{stabeq1}). Substituting \eqn{turning} and \eqn{zeromode} in the boundary equation (\ref{stabeq2}) the following transcendental equation must be satisfied
\bea
\label{transcendental}
{_2F_1}\left({3\over4},{3\over2},{7\over4};1-\l^2\right)={3\over2\l(1+\l^2)}\ .
\eea
Using \eqn{turning} and \eqn{transcendental} a critical value for $a$ is found numerically,
$a\simeq 0.813$\ , below which the
system becomes unstable. 

The conclusion of this analysis is that in the probe brane approximation non-singlet baryons with $0.813<k\leqslant N$ may exist for 
all Einstein internal manifolds bearing five-form flux. 
In the next
subsection we take the internal manifold to be Sasaki-Einstein and we switch on an instantonic magnetic flux proportional to the K\"ahler form. The $T^{1,1}$ and $S^5$ cases will be treated as particular examples, taking due care of the different periodicities.
The energy of the D5-brane will depend then on both the magnetic flux and the radius of $AdS$, and the same calculation above shows that non-singlet states exist as long as the number of quarks is larger than a minimum value that depends now on the volume of the $Y^{p,q}$. In fact the largest minimum value is reached for the $S^5$, contrary to our expectations that non-singlet baryons would be more restricted in less supersymmetric backgrounds. We review some basic facts about the geometry of $Y^{p,q}$ manifolds suitable for this study in the Appendix. 

\subsection{The baryon vertex in $AdS_5\times Y^{p,q}$ with magnetic flux}

Let us take the $AdS_5\times Y^{p,q}$ geometry and add a magnetic flux 
\begin{equation}
\label{theflux}
F={\cal N}J\, , 
\end{equation}
with $J$ the K\"ahler form 
of the 4 dimensional K\"ahler-Einstein submanifold of the $Y^{p,q}$, which solves the equations of motion.  As compared to the analysis in the previous subsection the presence of the magnetic flux will turn  the parametric region for which a classical solution exists to depend on $(a, \rho_0, {\cal N})$.

\noindent
For a non-vanishing ${\cal F}$ as above the energy of the D5-brane wrapped on the $Y^{p,q}$ is modified according to
\bea
\label{DBImacro}
E_{D5}={N\over 8\pi}\rho_0 \left(1+\frac{4\pi^2{\cal N}^2}{R^4}\right)
\eea
where we have used \eqn{j4}, and the fact that $J$ is self-dual and the determinant inside the square root is a perfect square. 

\noindent This magnetic flux dissolves irrational D1-brane charge in the $Y^{p,q}$, as inferred from the coupling
\begin{equation}
S_{D5}^{CS}=\frac12 (2\pi)^2 T_5\int_{\mathbb{R}\times Y^{p,q}}C_2\wedge F\wedge F=\frac{{\cal N}^2}{8} \frac{q^2[2p+(4p^2-3q^2)^{1/2}]}{p^2[3q^2-2p^2+p(4p^2-3q^2)^{1/2}]}\,T_1\int_{\mathbb{R}\times S^1_\psi}C_2
\end{equation}
This implies that the configuration will not allow a complementary description in terms of D1-branes expanding into a fuzzy 4 dimensional submanifold of the $Y^{p,q}$. 
We will see however that it will be possible to provide such a description in terms of D3-branes expanding into a fuzzy 2-sphere submanifold of the $Y^{p,q}$. In this case the magnetic flux that needs to be switched on will be proportional to the K\"ahler form on the $S^2$. We postpone this discussion to the Appendix.

In the $T^{1,1}$ case (see Appendix A.2 for a brief discussion of the $T^{1,1}$ geometry) our ansatz (\ref{theflux}) dissolves ${\cal N}^2/9$ D1-brane charge in the $T^{1,1}$, as implied by
\begin{equation}
S_{D5}^{CS}=\frac12 (2\pi)^2 T_5\int_{\mathbb{R}\times T^{1,1}}C_2\wedge F\wedge F=\frac{{\cal N}^2}{9}T_1\int_{\mathbb{R}\times S^1_\psi}C_2
\end{equation}
where we have used the second condition in (\ref{quantJ}). But in this case ${\cal N}^2/9$ is an integer due to Dirac quantization condition plus the first equation in (\ref{quantJ}). In this case a microscopical description in terms of expanding D1-branes will make sense, as we will show explicitly in section 6. 

Note that in fact for the $T^{1,1}$ we can take a more general ansatz for the magnetic flux, namely
$F={\cal N}_1\, J_1 \,+\, {\cal N}_2\, J_2$, with $J_1$, $J_2$ the K\"ahler forms on each of the $S^2$'s contained in the $T^{1,1}$. In this case the magnetic flux is dissolving ${\cal N}_1/3$ and ${\cal N}_2/3$ D3-brane charge in each $S^2$, and ${\cal N}_1 {\cal N}_2/9$ D1-brane charge in $S^2\times S^2$, as inferred from the couplings
\begin{equation}
S_{D5}^{CS}=2\pi\, T_5\int_{\mathbb{R}\times T^{1,1}} C_4\wedge F=\frac{{\cal N}_1}{3}\,T_2\int_{{\mathbb{R}\times S^1_\psi\times S_2^2}} C_4+
\frac{{\cal N}_2}{3}\,T_2\int_{{\mathbb{R}\times S^1_\psi\times S_2^1}} C_4
\end{equation}
and
\begin{equation}
S_{D5}^{CS}=\frac12 (2\pi)^2 T_5 \int_{\mathbb{R}\times T^{1,1}} C_2\wedge F\wedge F=\frac{{\cal N}_1 {\cal N}_2}{9}\, T_1\int_{\mathbb{R}\times S^1_\psi} C_2\ .
\end{equation}
Therefore ${\cal N}_1, {\cal N}_2\in 3\mathbb{Z}$, in agreement with Dirac quantization condition, as implied from (\ref{quantJ}). In this case the 
energy of the D5 is modified according to
\bea
\label{DBImacro2}
E_{D5}={N\over 8\pi}\rho_0 \sqrt{1+\frac{4\pi^2{\cal N}_1^2}{R^4}}\sqrt{1+\frac{4\pi^2{\cal N}_2^2}{R^4}}\ .
\eea

Coming back to the general case for $Y^{p,q}$ manifolds, $F={\cal N}J$, with $J$ the K\"ahler form of the 4 dimensional K\"ahler-Einstein submanifold of the $Y^{p,q}$,
 from \eqn{force} we find that the no-force condition on the $\rho$-axis yields
\bea
\label{turning2}
\rho_1=\rho_0(1-\l_{\rm eff}^2)^{1/4}\ ,\qq \l_{\rm
eff}={5-4a_{\rm eff}\over 4a_{\rm eff}}\ , 
\eea
where $a_{\rm eff}$ includes now the magnetic flux
\begin{equation}
a_{\rm eff}\equiv{a\over
1+{4\pi^2 {\cN^2}\over 5R^4}}\ .
\end{equation}
Given that $\l_{\rm eff}<1 $ a baryon
configuration exists for
\bea
a_{\rm eff} >  a_< \qquad {\rm with}\qquad
a_<={5\over8}+\frac{\pi^2{\cal N}^2}{2R^4} \, .
\eea
In terms of the volume of the $Y^{p,q}$ this reads
\begin{equation}
\label{avolume}
a_< = \frac58 + \frac{{\cal N}^2}{8\pi^2\, N}{\rm Vol}(Y^{p,q})\ ,
\end{equation}
so the bound depends now on the volume of the internal manifold. The largest volume given by the $Y^{p,q}$ metrics occurs for the $Y^{2,1}$, for which ${\rm Vol}(Y^{2,1})\approx 0.29 \pi^3$. Therefore we have that $\pi^3={\rm Vol}(S^5)>16/27\, \pi^3={\rm Vol}(T^{1,1})>{\rm Vol}(Y^{2,1})$ and $a_<$ is maximum for the $S^5$, the maximally supersymmetric case.
Note that since $a_{\rm eff} \leqslant 1$ there is also a bound on the instanton
number, namely 
\bea
a_< \leqslant 1\quad
\Rightarrow\quad \frac{{\cal N}^2}{R^4} \leqslant {3\over4\pi^2}\simeq 0.0761\ .
\eea

\noindent
Finally, the binding energy in terms of the physical length of the baryon reads
\bea
E=-{R^2\over2\pi L}{k\sqrt{1-\l_{\rm eff}^2}\over 3}\left({\cal J}-
{5-4a_{\rm eff}\over 4 a_{\rm eff}}\right){\cal I}\ .
\eea

\subsubsection{Stability analysis}

The study of the stability against longitudinal fluctuations gives again $\delta r(\rho)$ as in (\ref{zeromode}) where now ${_2F_1}(a,b,c;x)$ must satisfy \cite{SS2}
\bea
\label{transcendental2}
{_2F_1}\left({3\over4},{3\over2},{7\over4};1-\l_{\rm eff}^2\right)={3\over2\l_{\rm eff}(1+\l_{\rm eff}^2)}\ .
\eea
The critical value for $a_{\rm eff}$ that is found numerically is again
$a_{\rm eff}\simeq 0.813$\ , below which the
system becomes unstable. This improves the above bound for the
instanton number, in comparison to the 't Hooft coupling, to
\be
{{\cal N}^2\over R^4 } \lesssim 0.00291\ ,
\ee
that should be respected for the classical configuration not only to
exist, but also to be perturbatively stable. Thus, the stability
analysis sets a low bound for $a$ which is still less than unity.

\section{The baryon vertex in $\beta$-deformed backgrounds}

The description of the baryon vertex in these backgrounds is essentially identical to the one performed  in the previous section. 
Even though the $C_2$ and $B_2$ potentials are non-vanishing the tadpole introduced with the brane has still charge $N$, so it has 
to be compensated with the same number of fundamental strings attached. Moreover, the energy of the D5-brane wrapped on the 
deformed $S^5$ is the same as the one wrapped on the $S^5$. Therefore we find the same bound for the number of quarks that can form non-singlet baryons.

We discuss the general case of multi ${\hat \gamma}_i$-deformations \cite{Frolov}, from which the one- parameter Lunin-Maldacena background 
\cite{LM} is obtained for all ${\hat \gamma}_i$ equal. As we have mentioned, in this background the baryon vertex is described in terms 
of a D5-brane wrapped on the deformed $S^5$, with $N$ fundamental strings attached.
In the last subsection we will switch on a magnetic flux that on the one hand will increase the parametric space on which classical 
solutions exist and on the other hand will allow a complementary description in terms of expanding D3-branes suitable for the discussion 
of the finite 't Hooft coupling regime of the dual gauge theory. 

The multi-${\hat \gamma}_i$ deformed background reads in string frame \cite{Frolov}
\bea
\label{Frolov}
	ds^2 &=& R^2 \left[ds^2_{AdS_5} +  \sum_i \left(d \mu_i^2+ {\cal G} \mu_i^2 d\phi_i^2\right)
		+ {\cal G} \mu_1^2 \mu_2^2 \mu_3^2 \big( \sum_i \gh_i d\phi_i \big)^2
		\right]\,,\\\nn
	e^{2\phi} &=& {\cal G}\ ,\qq  {\cal G}^{-1} = 1 +\gh_3^2\,\mu_1^2 \mu_2^2 + \gh_1^2\,\mu_2^2 \mu_3^2 + \gh_2^2\,\mu_3^2 \mu_1^2\,. \\\nn
	B_2 &=& R^2 {\cal G}\, (\gh_3\,\mu_1^2 \mu_2^2 \,d\phi_1 \wedge d\phi_2
		+ \gh_1\,\mu_2^2 \mu_3^2\, d\phi_2 \wedge d\phi_3 + \gh_2\,\mu_3^2 \mu_1^2\, d\phi_3 \wedge d\phi_1)\,,\\\nn
	C_2 &=& - 4 R^2  \omega_1 \wedge (\gh_1 d\phi_1 +\gh_2 d\phi_2 +\gh_3 d\phi_3)\,,\\\nn
	C_4 &=& \omega_4 + 4 R^4 {\cal G}\, \omega_1\, \wedge d\phi_1 \wedge d\phi_2 \wedge d\phi_3\,,
\eea
where $\mu_i$ and $\phi_i$ parameterize a deformed five-sphere, so that we can write:
\begin{equation}\label{forms}
\begin{gathered}
	\mu_1 = \cos \alpha\,,\qquad
	\mu_2 = \sin \alpha \cos \theta\,,\qquad
	\mu_3 = \sin \alpha \sin \theta\,,\qquad
	\sum_{i=1}^3 \mu_i^2 = 1\,,\\
(\a,\th)\in[0,\pi/2]\ ,\qq
	d \omega_1 = \cos \alpha \sin^3 \alpha \sin \theta \cos \theta d \alpha \wedge d \theta\,,\qquad
	d \omega_4 = \omega_{AdS_5}\,.
\end{gathered}
\end{equation}
For equal ${\hat \gamma}_i$ parameters ${\hat \gamma}_1={\hat \gamma}_2={\hat \gamma}_3={\hat \gamma}$, ${\hat \gamma}$ is related 
to the deformation parameter $\beta$ of the gauge theory through \cite{LS}:
\begin{equation}
	\gh = R^2\ \beta\,.
\end{equation}

\subsection{The D5-brane baryon vertex}

Let us now consider a D5-brane wrapping the deformed $S^5$ in (\ref{Frolov}). This brane captures the $F_5-F_3\wedge B_2$ flux of the background but still requires $N$ fundamental strings to cancel the tadpole
\begin{eqnarray}
\label{tadpole2}
S_{D5}^{CS}&=&2\pi\, T_5\int_{\mathbb{R}\times {\tilde S}^5} P\left[C_4-C_2\wedge B_2\right]\wedge F=-2\pi\, T_5 \int_{\mathbb{R}\times {\tilde S}^5} P\left[F_5-F_3\wedge B_2\right]\wedge A
=\nonumber\\
&=&-N \int dt A_t
\end{eqnarray}
since
\bea
\label{CS.beta}
F_5-F_3\wedge B_2=%\o_4+4R^4 G\o_1\wedge d\phi_1\wedge d\phi_2\wedge d\phi_3\\ \nn
%&+&4 R^2\gh\o_1\wedge(d\phi_1+d\phi_2+d\phi_3)
%\wedge R^2\gh G(\m_1^2\m_2^2 d\phi_1\wedge d\phi_2+\m_1^2\m_3^2 d\phi_3\wedge d\phi_1+\m_2^2\m_3^2 d\phi_2\wedge d\phi_3)
\omega_{AdS_5}+4 R^4\, d\omega_1\wedge d\phi_1\wedge d\phi_2\wedge d\phi_3~,
\eea
as in the $AdS_5\times S^5$ case. Therefore the CS part of the action is undeformed.
The DBI action is in turn given by
\begin{equation}
\label{D5wv}
	S_{D5}^{DBI} =- T_5 \int_{\mathbb{R}\times\tilde{S}^5} d^{6}\xi \ e^{-\phi}
		\sqrt{- \det P\left(G + 2\pi F - B_2\right) }\, .
\end{equation}
For $F=0$ the determinant of the pull-back of $G-B_2$ can be written as
\bea
\label{det}
\D_1=G_{tt}G_{\a\a}G_{\th\th}  \det{\G}~,
\eea
where $\G$ is a $3\times3$ matrix  of the form $G_{ab}-B_{ab}$ for the three $U(1)$ directions. We then get
\bea
\label{det1}
\D_1
%=\frac{G_{tt} R^{10} \cos \a^2 \cos\th^2 \sin\a^6 \sin\th^2}{1+\gh^2 \sin\a^2 \left(\cos \a^2+\cos\th^2 \sin\a^2 \sin\th^2\right)}
=G_{tt}\, R^{10} \sin^6{\a} \cos^2{\a} \sin^2{\th} \cos^2{\th}\, {\cal G}\ ,
\eea
such that the DBI action remains also undeformed.

Given that the $AdS$ part of the background remains untouched by the deformation the contribution of the fundamental strings stretching 
from the D5 to the boundary of $AdS$ is the same as in the undeformed case. The only issue here would be that the binding energy was 
modified due to the dependence of the D5-brane on the deformation parameter. We have shown however that this dependence drops out 
both in the CS and DBI actions. Therefore the size and binding energy of the baryon remain undeformed, and coincide with those 
in ${\cal N}=4$. The classical solution and its stability analysis are therefore identical to those performed in section 3. Last but not least,
we should mention that for marginally-deformed backgrounds there are cases on which the classical solution coincides with $\cN=4$,
does not depend on the deformation parameter, but the stability analysis even for the conformal case requires an upper value on the imaginary part of the deformation
parameter $\s$ as in the case of mesons \cite{Avramis:2007mv}.

\subsubsection{Adding a magnetic flux}

Finally we can switch on a magnetic flux $F={\cal N}J$, with ${\cal N}\in {\mathbb{Z}}/2$ and $J$ the K\"ahler form of the $S^2$ parameterized by $(\alpha,\theta)$, dissolving $2{\cal N}$ units of D3-brane charge in the baryon. This will allow a microscopical description in terms of expanding D3-branes from which the finite 't Hooft coupling region can be studied. The DBI action changes as
\begin{equation}
\label{D5deformed}
S^{DBI}_{D5}=-\frac{T N}{8\pi}\rho_0\sqrt{1+\frac{4\pi^2{\cal N}^2}{R^4}}\ .
\end{equation}
In the presence of a magnetic flux the minimum number of quarks forming a non-singlet baryon  is modified and there is a maximum on the magnetic 
flux that can be dissolved in the baryon, in parallel with what we have found in the previous section.

\section{The baryon vertex in the Maldacena--Nu\~nez background}

The Maldacena-Nu\~nez background \cite{MN} is a solution to Type IIB supergravity dual to a ${\cal N}=1$ supersymmetric confining gauge theory. It can be obtained as a solution of seven dimensional gauged supergravity \cite{CV}, uplifted to ten dimensions. Given that this background is confining we expect that the universality of the baryon vertex configurations found in the previous conformal examples (in the absence of a magnetic flux) is lost. This is indeed confirmed by the analysis in this section. 

\subsection{The Maldacena--Nu\~nez background}

%The background and includes the metric the dilaton and the RR three-form.
The ten-dimensional metric  reads in the string frame 
\bea
ds^2_{10}\,=\,e^{\phi}\,\,\Big[\,
dx^2_{1,3}\,+g_sN\Bigl(\,e^{2h}\,\big(\,d\theta_1^2+\sin^2\theta_1 d\phi_1^2\,\big)\,+\,
d\rho^2\,+\,{1\over 4}\,(w^i-A^i)^2\Bigr)\,\Big]\,\,,
\label{metricMN}
\eea
where $\phi$ is the dilaton, $h$ is a function of the radial coordinate $\rho$, the
one-forms $A^i$ $(i=1,2,3)$ are the components of
the non-abelian gauge vector field of the seven-dimensional gauged supergravity,
\bea
A^1\,=\,-a(\rho) d\theta_1\,,
\,\,\,\,\,\,\,\,\,
A^2\,=\,a(\rho) \sin\theta_1 d\phi_1\,,
\,\,\,\,\,\,\,\,\,
A^3\,=\,- \cos\theta_1 d\phi_1\,,
\label{oneform}
\eea
and the $w^i$'s are the right-invariant Maurer-Cartan dreibeins of $SU(2)$,
satisfying  $dw^i=-{1\over 2}\,\varepsilon_{ijk}\,w^j\wedge w^k$. They
define a three-sphere that can be parameterized as
\bea
w^1&=& \cos\psi\, d\theta_2\,+\,\sin\psi\sin\theta_2\,
d\phi_2\,\,,\\
w^2&=&-\sin\psi\, d\theta_2\,+\,\cos\psi\sin\theta_2\,
d\phi_2\,\,,\nonumber\\
w^3&=&d\psi\,+\,\cos\theta_2\, d\phi_2\,\,.\nonumber
\eea
The angles $\theta_\a,\phi_\a\ ,\a=1,2$ and $\psi$ take values in the intervals $\theta_i\in [0,\pi]$,
$\phi_i\in [0,2\pi]$ and $\psi\in [0,4\pi]$. The functions $a(\rho)$, $h(\rho)$ and the dilaton $\phi(\rho)$
are given by
\bea
&&a(\rho)={2\rho\over \sinh 2\rho}\ ,\qq e^{2h}=\rho\coth 2\rho\,-\,{\rho^2\over \sinh^2 2\rho}\,-\,
{1\over 4}\ , \quad\\
&&e^{2\phi}=e^{-2\phi_0}\,{\sinh 2\rho\over 2e^h}\equiv e^{-2\phi_0}\L(\rho)\ ,\qq e^{2\phi_0}=g_sN\ .
\eea
In particular, $\Lambda(\rho)$ satisfies
\begin{equation}
\L(\rho)\simeq\frac{e^{2\rho}}{4\sqrt{\rho}}\ , \quad {\rm when} \quad \rho\gg1
\end{equation}
and
\begin{equation}
\L(\rho)\simeq1+\frac{8\rho^2}{9}+\cO(\rho^4)\ ,\quad {\rm when} \quad  \rho\ll1\ .
\label{MNsol}
\end{equation}
\no
The solution also includes a Ramond-Ramond three-form 
given by
\bea
F_3\,=\,{g_sN\over 4}\left\{-\,\big(\,w^1-A^1\,\big)\wedge
\big(\,w^2-A^2\,\big)\wedge \big(\,w^3-A^3\,\big)\,+\,\,\,
\sum_i\,F^i\wedge \big(\,w^i-A^i\,\big)\right\}\,\,,
\label{RR3}
\eea
where $F^i$ is the field strength of the $SU(2)$ gauge field $A^i$, defined as
$F^i\,\equiv\,dA^i\,+\,{1\over 2}\varepsilon_{ijk}\,A^j\wedge A^k$.

\subsection{The D3-brane baryon vertex}

A D3-brane wrapping the 3-sphere parameterized by 
$(\th_2, \phi_2, \psi)$ introduces a tadpole that needs to be canceled through the addition of $N$ fundamental strings
\bea
\label{CS.MN}
S_{D3}^{CS}=2\pi \,T_3\int_{\mathbb{R}\times S^3} C_2\wedge F=-2\pi\,T_3\int_{\mathbb{R}\times S^3} F_3\wedge A=-N\int dt\,  A_t\ . 
\eea
The DBI action of this D3-brane is given by:
\bea
\label{DBI.MN}
S_{D3}^{DBI}&=&-T_3\int_{\mathbb{R}\times S^3} d^4\xi \, e^{-\phi}\sqrt{-\det P(G)}=
-\frac{TN}{4\pi}\sqrt{\Lambda(\rho_0)}\ .
\eea
Particularizing to this background the size of the vertex given by (\ref{length}) we find
\bea
\label{Length.MN}
L=\sqrt{g_sN}\int_{\rho_0}^\infty\frac{d\rho}{\sqrt{\L(\rho)/\L(\rho_1)-1}}\ ,
\eea
which is a decreasing function of $\rho_0$. 
The binding energy of the baryon is in turn given by 
\begin{equation}
E=\frac{k}{2\pi}\Bigl\{\int_{\rho_0}^\infty d\rho \frac{\Lambda(\rho)}{\sqrt{\Lambda(\rho)-\Lambda(\rho_1)}}-\int_{\rho_{\rm min}}^\infty d\rho\, 
\sqrt{\Lambda(\rho)}+\frac{1-a}{a}\int_{\rho_{\rm min}}^{\rho_0} d\rho\, \sqrt{\Lambda(\rho)}+\frac{1}{2a}\sqrt{\Lambda(\rho_0)}\Bigr\}\ .
\end{equation}
Both integrals receive most of their contributions from the region $\rho\approx \rho_1$ so it can be seen that $E$ is linearly proportional to $L$ \cite{Loewy:2001pq}. Also, from (\ref{Evar}) we see that $E$ and $L$ share the same dependence on the position of the vertex:
\begin{equation}
\frac{dE}{d\rho_0}=\frac{k\sqrt{\L_1}}{2\pi\sqrt{g_sN}}\frac{dL}{d\rho_0}\, .
\end{equation}
The net-force condition is now
\bea
\label{force.MN}
\cos\Th=\frac{1-a}{a}+\frac{1}{4a}\del_\rho\ln \L(\rho)\bigg{|}_{\rho=\rho_0}\ ,\qq \cos\Th=\sqrt{1-\frac{\Lambda_1}{\Lambda_0}}\ .
\eea
Taking into account that $\del_\rho\ln \L(\rho)$ satisfies $\del_\rho\ln \L(\rho)\lesssim2-1/(2\rho)+\cO(1/\rho^2)$ in the UV 
we find that $a>a_{<}$ with $a_{<}=3/4$. Therefore the minimum value of the number of quarks is restricted with respect to the one found in the previous conformal examples, in agreement with our expectations.

\subsubsection{Stability analysis}

The study of the stability against longitudinal fluctuations gives
\bea
\label{Stability}
\delta r(\rho)= A\, g_sN\int_{\rho}^\infty d\rho\, \frac{\L}{(\L-\L_1)^{3/2}}\ , 
\eea
as the solution to equation (\ref{stabeq1}). Substituting in the boundary equation (\ref{stabeq2})
we find
\bea
2(\L-\L_1)\delta r^\prime+\Lambda^\prime (\rho)\, \delta r=0 \quad {\rm at} \quad \rho=\rho_0\, ,
\eea
and using \eqn{force.MN} we can write
\bea
\label{zero.MN}
a\,(\cos\Th+\frac{a-1}{a})\cos\Th\ \cZ=\frac12
\eea
where
\bea
\cZ\equiv \sqrt{\L_0}\int_{\rho_0}^\infty d\rho\,\frac{\L}{(\L-\L_1)^{3/2}}\ ,\quad {\rm and} \quad \L_1=\L_0\sin^2\Th\ .
\eea
From (\ref{zero.MN}) we can now solve for $a$. Note that using 
\eqn{derivative} we find that
\bea
\label{derivativeMN}
\frac{\partial \rho_0}{\partial \rho_1}=\half\cZ\cos\Th\ \del_{\rho}\ln\L(\rho)|_{\rho=\rho_1}
\eea
from where
\bea
\cZ\cos\Th=\frac{2}{\del_{\rho_0}\ln\L(\rho_0)}\in[1,\infty)\ .
\eea
From \eqn{zero.MN} and \eqn{derivativeMN} we then find
$\displaystyle a\geqslant\half+\frac{1}{4\cZ\cos\th}\Longrightarrow a\geqslant\half+\frac{\del_{\rho_0}\ln\L(\rho_0)}{8}$.
Thus, the stability analysis does not improve the bound imposed by the existence of a classical solution, in contrast to what happened in the conformal examples previously discussed. 

\subsubsection{Adding a magnetic flux}

Finally, in order to compare with the microscopical analysis in section 6.3 we add a magnetic flux to the baryon proportional to the K\"ahler form on the 2-sphere parameterized by $(\theta_2, \phi_2)$, $F={\cal N}J$, with ${\cal N}\in 2 \mathbb{Z}$. This flux dissolves ${\cal N}/2$ units of D1-brane charge in the $S^3$. The energy of the baryon is modified according to 
\begin{equation}
\label{D3macro}
E_{D3}=\frac{N}{4\pi}\sqrt{\Lambda(\rho_0)+\frac{4\pi^2{\cal N}^2}{g_s N}}\ .
\end{equation}
As in the previous cases the magnetic flux changes the minimum bound for the number of quarks in the baryon.  Moreover the flux has an upper bound.

\section{The microscopical description}

In the previous sections we have discussed generalizations of the baryon vertex constructions to allow a magnetic flux dissolving lower dimensional brane charge in the configuration. By analogy with Myers dielectric effect \cite{Emparan, Myers} we expect that a complementary description in terms of  lower dimensional branes expanding into fuzzy baryons should then be possible. This would be the ``microscopical'' realization of the ``macroscopical'' baryons with magnetic flux that we have just described. The interesting thing about the microscopical description is that it allows to explore the finite 't Hooft coupling region, and this is especially relevant in those cases in which the baryons are non-supersymmetric, like those considered in this paper, and are therefore not preserved by a BPS condition.

It is well known that the macroscopical and microscopical descriptions have complementary ranges of validity \cite{Myers}. While the first is valid in the supergravity limit the second is a good description when the mutual separation of the expanding branes is much smaller than the string length, such that they can be taken to be coincident and therefore described by the $U(n)$ effective action constructed by Myers \cite{Myers}.
For $n$ D$q$-branes expanded into an $r$-dimensional manifold of radius $R$, the volume per brane can be estimated as $R^{r-q}/n$, which must then be much smaller than $l_s^{r-q}$. Thus the condition
\begin{equation}
\label{microreg}
R<<n^{\frac{1}{r-q}}\, l_s\, ,
\end{equation}
sets the regime of validity of the microscopical description.  The macroscopical description is in turn valid when $R>>1$. Therefore both descriptions are complementary for finite $n$, but should agree in the large $n$ limit, where they have a common range of validity. The limit (\ref{microreg}) is especially appealing in backgrounds with a CFT dual, like the AdS spacetimes that we have considered in this paper. Indeed, in terms of the 't Hooft parameter of the dual CFT the condition (\ref{microreg}) reads 
\begin{equation}
\label{microreg2}
\lambda<<n^{\frac{4}{r-q}}\, .
\end{equation}
The fact that $\lambda$ can be finite opens up the possibility of accessing the finite 't Hooft coupling region of the dual CFT through the microscopical study of the corresponding dual brane system. 

Dielectric branes expanding into fuzzy manifolds have been extensively studied in the literature. From (\ref{microreg2}) the lower the dimensionality of 
the expanding branes the smaller the 't Hooft parameter can get. However for the manifolds that we have discussed in this paper it will not always be 
possible to provide a description in terms of expanding D1-branes. This is the case for the $Y^{p,q}$ Sasaki-Einstein geometries, in which the natural 
microscopical description would be in terms of D1-branes wrapped on the Reeb vector direction and expanding into the remaining four dimensional K\"ahler-Einstein 
manifold. We are however not aware of a fuzzy realization of these manifolds besides the $CP^2$ case. Moreover, as we have seen, the number of D1-branes 
in the macroscopical description is irrational, while this should be an integer in the microscopical description.
Still, we will be able to provide a (less) microscopical description in terms of D3-branes expanding into a fuzzy 2-sphere. For 
the ${\hat \gamma}_i$ deformed backgrounds the natural thing is to dissolve D3-branes wrapped on the $(T^1)^3$ through the addition 
of a magnetic flux proportional to the K\"ahler form on the $S^2$, as we did in section 4.1. The microscopical description will then 
be in terms of D3-branes expanding into a fuzzy $S^2$.

We start in section 6.1 with the analysis of the $AdS_5\times T^{1,1}$ background, for which a description in terms of D1-branes expanding into a fuzzy $S^2\times S^2$ manifold can be done. As we will see this description exactly matches the macroscopical description in section 3.2. The extension to arbitrary $Y^{p,q}$ manifolds is more technical and it is postponed to the Appendix. In section 6.2 we discuss the ${\hat \gamma}_i$ deformed backgrounds. We end with the Maldacena-Nu\~nez analysis in section 6.3, in terms of D1-branes expanding into a fuzzy $S^2$ baryon.

\subsection{The $AdS_5\times T^{1,1}$ background: D1-branes into fuzzy $S^2\times S^2$}

The DBI action describing the dynamics of $n$ coincident D1-branes is given by \cite{Myers}
\begin{equation}
\label{D1myers}
S^{DBI}_{nD1}=-T_1\int d^2\xi \, {\rm STr}\Bigl\{e^{-\phi} \sqrt{|{\rm det}
\Bigl(P[E_{\mu\nu}+E_{\mu i}(Q^{-1}-\delta)^i{}_j
E^{jk}E_{k\nu}]\Bigr){\rm det}Q|}\Bigr\}
\end{equation}
where $E=G-B_2$ and
\begin{equation}
\label{Qmatrix}
Q^i{}_j=\delta^i{}_j+\frac{i}{2\pi}[X^i,X^k]E_{kj}\, .
\end{equation}
Let us take the D1-branes wrapped on the $U(1)$ fibre direction $\psi$ in (\ref{T11}) and  expanding into the fuzzy $S^2\times S^2$ submanifold parameterized by $(\theta, \phi)$ and $(\omega, \nu)$. 

Using Cartesian coordinates for each $S^2$ we can impose the condition
\begin{equation}
\sum_{i=1}^3 (x^i)^2=1
\end{equation} 
at the level of matrices if the $X^i$ are taken in the irreducible totally symmetric representation of order $m$, with dimension $n=m+1$, 
\begin{equation}
\label{noncom}
X^i=\frac{1}{\sqrt{m(m+2)}}\, J^i
\end{equation}
with $J^i$ the generators of $SU(2)$, satisfying $[J^i, J^j]=2i\varepsilon_{ijk} J^k$. Labeling with $m_1$, $m_2$ the irreps 
for each $S^2$ we have that the total number of expanding branes $n=(m_1+1)(m_2+1)$, and
substituting in the DBI action
\bea
\label{DBImicroKW}
S_{n D1}^{DBI}=-T_1\int d^2\xi \sqrt{-G_{tt}G_{\psi\psi}}\ \rm{Str}\sqrt{\det Q}
\eea
we find
\bea
E_{n D1}= \frac{N\rho_0}{8\pi} \frac{(m_1+1)(m_2+1)}{\sqrt{m_1(m_1+2)m_2(m_2+2)}} \sqrt{1+\frac{36\pi^2 m_1(m_1+2)}{R^4}}\sqrt{1+\frac{36\pi^2 m_2(m_2+2)}{R^4}}
\eea
where 
\begin{equation}
{\rm det}\, Q=\Bigl(1+\frac{R^4}{36\pi^2 m_1(m_1+2)}\Bigr)\Bigl(1+\frac{R^4}{36\pi^2 m_2(m_2+2)}\Bigr)\mathbb{I}
\end{equation}
and the $(m_1+1)(m_2+1)$ factor comes from computing the symmetrized trace. This expression is exact in the limit
\begin{equation}
\label{limit}
R>>1\, , \qquad m>>1\, , \qquad {\rm with} \qquad \frac{R^2}{m}={\rm finite}
\end{equation}
(see section 5.1 of \cite{Lozano:2011} for the detailed discussion).
Taking the large $m_1$, $m_2$ limit  we find perfect match with the macroscopical result given by (\ref{DBImacro2}) if $m_1\sim {\cal N}_1/3$, $m_2\sim {\cal N}_2/3$,
in agreement with (\ref{quantJ}).

\subsubsection{The F-strings in the microscopical description}

An essential part of the baryon vertex are the fundamental strings that stretch from the D$p$-brane to the boundary of $AdS_5$. As we show in this section they arise from the non-Abelian CS action.

The CS action for $n$ coincident D1-branes is given by
\begin{equation}
\label{CSaction}
S_{CS}=\int d^2\xi\, {\rm STr}
\Bigl\{ P \Bigl(e^{\frac{i}{2\pi}(i_X i_X)}\, \sum_q C_q \,\, e^{-B_2}\Bigr) e^{2\pi F} \Bigr\}\ .
\end{equation}
In this expression the dependence of the background potentials on the
 non-Abelian scalars occurs through the Taylor expansion
\cite{GM}
\begin{equation}
\label{Taylorex}
C_q(\xi,X)=C_q(\xi)+X^k \partial_k C_q(\xi)+\frac12 X^l X^k \partial_l \partial_k C_q (\xi)+\dots
\end{equation}
and it is implicit that the pull-backs into the worldline are taken with gauge covariant derivatives
$D_\xi X^\mu=\partial_\xi X^\mu+i[A_\xi, X^\mu]$.

The relevant CS couplings in the $AdS_5\times T^{1,1}$ background are
\bea
\label{CSmicro}
S_{n D1}^{CS}=\frac{T_1}{2\pi}\int d^2\xi\, {\rm Str}\left(i P[(i_Xi_X)C_4]-\frac12 P[(i_Xi_X)^2C_4] \wedge F\right)\ .
\eea
Taking into account (\ref{Taylorex}) and working in the gauge $A_\psi=0$ these couplings reduce to
\begin{equation}
S^{CS}_{n D1}=-\frac{1}{\pi}\int dt\, {\rm Str}\Bigl[(i_X i_X)^2 i_k F_5\Bigr]  A_t\ ,
\end{equation}
where $i_k$ denotes the interior product with $k^\mu=\delta^\mu_\psi$ and we have integrated out $\psi$, the spatial direction of the D1-branes.
Taking into account that in Cartesian coordinates $F_5$, as given by (\ref{F5T11}), reduces to
\begin{equation}
i_k F_5=-\frac{R^4}{27} f_{ijm}f_{kln} X^m X^n dX^i \wedge dX^j \wedge dX^k \wedge dX^l\, ,
\end{equation}
where the indices run from 1 to 3 for the first 2-sphere and from 4 to 6 for the second, such that  
$f_{ijm}=\varepsilon_{ijm}$ for $i,j,m=1,\dots 3$ and $i,j,m=4,\dots, 6$ and zero otherwise,
we finally find
\begin{equation}
S_{n D1}^{CS}=-N \frac{(m_1+1)(m_2+1)}{\sqrt{m_1(m_1+2)m_2(m_2+2)}}\int dt\, A_t\, ,
\end{equation}
again in perfect agreement with (\ref{tadpole}) in the large $m_1, m_2$ limit.

To finish this section we would like to point out that more general fuzzy realizations of the $T^{1,1}$ could in principle be considered. For instance one could think of substituting the direct product of the two fuzzy 2-spheres by a Moyal-type of product,  $[X^i,X^j]=\imath\theta^{ij}$ where $i=1,\dots 3$ refers to the first 2-sphere and $j=4,\dots 6$ refers to the second. It is not clear in any case how this would affect the description of the vertex beyond the supergravity limit.

\subsection{The $\beta$-deformed backgrounds: D3-branes into fuzzy $S^2$}

In this case we start with a system of $n$ coincident D3-branes, whose dynamics is given by the straightforward extension of (\ref{D1myers}) to a four dimensional worldvolume. We take the branes wrapped on the 3-torus and expanding into the 2-sphere in (\ref{Frolov}) parameterized by $(\alpha, \theta)$. Given that the expansion is on a fuzzy 2-sphere we take the same ansatz
(\ref{noncom}) as in the previous section. Substituting in the DBI action we have
\bea
\label{DBImicroLM}
&&S_{n D3}^{DBI}=-T_3\int d^4\xi\ \rm{Str}\left[e^{-\phi}\sqrt{-\det P[E_{\mu\nu}] \det Q}\right]
\eea
with
\begin{equation}
{\rm det}\, Q=\Bigl(1+\frac{R^4}{\pi^2 m(m+2)}\Bigr)\mathbb{I}
\end{equation}
as in the previous section for each 2-sphere\footnote{The different factor comes from the different radii in the two backgrounds.}. As explained there this expression is exact in the limit (\ref{limit}).
The only difference with the calculation in the previous section comes from the fact that ${\rm det}\, P[E_{\mu\nu}]$ depends now on the transverse scalars $X^i$, and therefore it contributes to the symmetrized trace. In order to compute this contribution we use that 
\bea
\label{abe}
\rm{Str}(\mu_1\mu_2\mu_3)\simeq\frac{m+1}{4\pi}\int_0^{\pi/2}d\a\sin^3\a\cos\a\int_0^{\pi/2}d\th\sin\th\cos\th=\frac{m+1}{32\pi} 
\eea
as implied by equation (4.44) in \cite{Abe}, which is valid in the limit $m\gg 1$. 
We then find that
\bea
\label{energyD3mic}
E_{n D3}=\frac{N\rho_0}{8\pi}\frac{m+1}{\sqrt{m(m+2)}}\sqrt{1+\frac{ \pi^2m(m+2)}{R^4}}\, .
\eea
This result for the energy is more approximate than the ones found in the rest of examples, where 
${\rm det}\, P[E_{\mu\nu}]$ does not depend on the transverse scalars. Still, it allows to compute $1/m$ corrections to the macroscopical result.
Taking the large $m$ limit we find perfect agreement with the macroscopical result (\ref{D5deformed}) for $m\sim 2{\cal N}$.

\subsubsection{The F-strings}

The relevant CS couplings in the non-Abelian action for D3-branes in the $\hat{\gamma}_i$ deformed backgrounds are
\begin{equation}
S_{n D3}^{CS}=T_3 \int d^4\xi\, {\rm STr}\Bigl(P[C_4]+i P[(i_X i_X) C_4]\wedge F-P[C_2\wedge B_2]-i P[(i_X i_X)(C_2\wedge B_2)]\wedge F\Bigr)\ .
\end{equation}
Using (\ref{Taylorex}) and the definition of the gauge covariant pull-backs they reduce to
\begin{equation}
S_{n D3}^{CS}= i\, T_3 \int d^4\xi\, {\rm STr}\Bigl({\cal P}[(i_X i_X)F_5]-{\cal P}[(i_X i_X)F_3\wedge B_2]\Bigr)A_t\ ,
\end{equation}
where ${\cal P}$ denotes the gauge covariant pull-back over the spatial directions. Taking the spatial components of the gauge field to vanish and using that
\begin{equation}
F_5=\frac{1}{8\pi} R^4\,  {\cal G}\, \varepsilon_{ijk}X^k dX^i \wedge dX^j\wedge d\phi_1\wedge d\phi_2\wedge d\phi_3
\end{equation}
and
\begin{equation}
F_2=-\frac{1}{8\pi} R^2 \varepsilon_{ijk}X^k dX^i \wedge dX^j\wedge ({\hat \gamma}_1 d\phi_1+ {\hat \gamma}_2 d\phi_2+{\hat \gamma}_3 d\phi_3)\, ,
\end{equation}
the $F_5$ and $F_3\wedge B_2$ contributions combine to give 
\begin{equation}
S_{n D3}^{CS}=-N \frac{m+1}{\sqrt{m(m+2)}}\int dt A_t
\end{equation}
which is again in perfect agreement with (\ref{tadpole2}) in the large $m$ limit.

\subsection{The Maldacena--Nu\~nez background: D1-branes into fuzzy $S^2$}

Let us now use the action (\ref{D1myers}) to describe $n$ D1-branes wrapped on the $\psi$ direction and expanding into the 2-sphere in (\ref{metricMN}) parameterized by $(\theta_2, \phi_2)$. The expansion is again on a fuzzy 2-sphere, so we take the same non-commutative ansatz (\ref{noncom}) as in the previous sections. Substituting in the DBI action we have
\bea
\label{DBImicroMN}
S_{n D1}^{DBI}=-T_1\int d^2\xi \sqrt{-G_{tt}G_{\psi\psi}}\ \rm{Str}\sqrt{\det Q}
\eea
as in (\ref{DBImicroKW}), with 
\begin{equation}
{\rm det}\, Q=\Bigl(1+\frac{g_s N \Lambda(\rho_0)}{16\pi^2 m(m+2)}\Bigr)\mathbb{I}\, .
\end{equation}
The regime of validity of the determinant is again fixed by (\ref{limit}). Computing the symmetrized trace we finally arrive at
\bea
\label{EnergymicroMN}
E_{n D1}=\frac{N}{4\pi}\frac{m+1}{\sqrt{m(m+2)}}\sqrt{\Lambda(\rho_0)+\frac{16\pi^2m(m+2)}{g_sN}}\ ,
\eea
which in the large $m$ limit is in perfect agreement with the macroscopical result (\ref{D3macro}) for $m\sim {\cal N}/2$.

\subsubsection{The F-strings}

The relevant CS couplings are in this case
\bea
S_{n D1}^{CS}=T_1\int{\rm Str}\, \Bigl(P[C_2]+iP[(i_Xi_X)C_2]\wedge F\Bigr)
\eea
which can be rewritten as
\bea
S_{n D1}^{CS}=2i \int dt \, {\rm STr}\Bigl[(i_X i_X)i_k F_3\Bigr] A_t
\eea
where $i_k$ denotes the interior product with $k^\mu=\delta^\mu_\psi$ and we have integrated over the $\psi$ direction. Using that
\begin{equation}
F_3=-\frac{N}{4} \varepsilon_{ijk} X^m dX^i \wedge dX^j \wedge d\psi
\end{equation}
we get
\begin{equation}
S_{n D1}^{CS}=-N \frac{m+1}{\sqrt{m(m+2)}}\int dt A_t
\end{equation}
in perfect agreement with (\ref{CS.MN}) in the large $m$ limit.

\vspace{0.5cm}

The analysis performed in this section shows that the right description for the baryon vertex (with magnetic flux) at finite 't Hooft coupling is in terms of D1- or D3-branes expanding into a $S^1\times (S^2\times S^2)_{\rm fuzzy}$ D5-brane, $S^2_{\rm fuzzy}\times T^3$ D5-brane or $S^1\times S^2_{\rm fuzzy}$ D3-brane. As we have shown these branes introduce tadpoles that need to be cancelled with the addition of fundamental strings. A full description of the D5, or D3, plus F1 system valid at finite 't Hooft coupling would require however the construction of fuzzy spikes, so that the $\alpha^\prime$ corrections coming from the F-strings would also be taken into account. See the conclusions for a further discussion on this point.

%\section{Finite 't Hooft coupling-Review at the end of typing}

%It is straightforward to prove that the exact expressions
%for the DBI and the CS actions read
%\bea
%\label{DBI.CS.finite}
%&&S^{nD1}_{DBI}=-\frac{\sqrt{-G_{tt}} RNT}{8\pi}\frac{(m+1)^2}{m(m+2)}
%\left(1+\frac{36\pi^2m(m+2)}{R^4}-\frac{\imath R^2}{9\pi m(m+2)}\right)\ ,\\
%&&S^{nD1}_{CS}=-N\frac{(m+1)^2}{m(m+2)}\left(1-\frac{14}{3m(m+2)}\right)\int A\equiv
%-rN\int A\ , \nonumber
%\eea
%where the range of validity of Myers construction is $(m+1)^2\gg 16\pi^2 R^4$. The exact expressions of the
%DBI and the CS action can be used in order to exploit the finite 't Hooft coupling corrections of the baryon energy.
%However, we have to consider the $\a^\prime$-corrections of the Nambu--Goto action and background which we leave it for future study.

\section{Conclusions}

In this paper we have discussed non-singlet baryon vertices in various Type IIB backgrounds in order to investigate the dependence of the bound imposed on the number of quarks by the existence and stability of the classical solution, on the supersymmetry and confinement properties of the dual gauge theory.

Using the probe brane approximation \cite{Brandhuber:1998, Imamura1, Loewy:2001pq} we have shown that this bound is the same for all 
$AdS_5 \times Y_5$ backgrounds with $Y_5$ an Einstein manifold bearing five form flux, independently on the number of supersymmetries preserved. 
The same result holds true for $\beta$-deformed and even non-supersymmetric multi-$\beta$ deformed backgrounds, pointing at a universal behavior
 based on conformality. The same analysis in a confining background, the Maldacena-Nu\~nez model, shows that universality is lost when confinement 
is present. In this case although non-singlet baryons still exist, the bound imposed on them is more restrictive, in agreement with our expectations 
that non-singlet baryons should be more constrained in more realistic gauge theories. It would be interesting to confirm this result in other 
confining backgrounds, such as the Klebanov-Strassler \cite{KS} or the Sakai-Sugimoto models \cite{Sakai:2004cn}.

Although the probe brane analysis has proved to be enough in order to deduce the basic properties of this type of systems (see for instance \cite{Malda, Brandhuber:1998, Imamura1, Loewy:2001pq}), the fact that all supersymmetries are broken in this approach could imply that it  may not be sensitive enough to account for the supersymmetries preserved by the different backgrounds. However previous results in the literature on baryon vertices in $AdS_5 \times T^{1,1}$, $AdS_5\times Y^{p,q}$ and the Klebanov-Strassler and Maldacena-Nu\~nez backgrounds reveal that even when all fundamental strings are taken to end on the same point of the wrapped D-brane supersymmetry is broken. Therefore significant changes to the probe brane results should not be expected. At any event,
the different behaviors based on conformality should represent valid predictions. 
 
We also note that we would expect the baryon analysis in $\beta$-deformed Sasaki--Einstein manifolds to provide similar results to the undeformed case. Our string and brane configurations do not seem to depend strongly on the deformation in the way encountered in \cite{gtbeta}, where important modifications due to the deformation appeared only in the $T^3$ fibration description.
 
Using the fact that we can consistently add lower dimensional brane charges we have provided an alternative description of the baryons in terms of lower dimensional branes expanding into fuzzy baryon vertices. This description represents a first step towards the analysis of holographic baryons at finite 't Hooft coupling. In this description the expansion is caused by a purely gravitational dielectric effect, while the Chern-Simons terms only indicate the need to introduce the number of fundamental strings required to cancel the tadpole. 

In order to be able to conclude that non-singlet baryons exist at finite 't Hooft coupling we should take into account not only the $\alpha^\prime$ corrections coming from the microscopical analysis of the brane but also the $\alpha^\prime$ corrections to the F-string Nambu-Goto action and the background. This is therefore a difficult program, which we have only begun to explore.
An interesting next step in this direction would be to use the microscopical analysis to build up spike solutions in these backgrounds. We expect to report progress in this direction in the near future.

\subsection*{Acknowledgements}

We would like to thank A. Chatzistavrakidis, N. Obers, A.V. Ramallo, D. Rodr\'{\i}guez-G\'omez and K. Sfetsos  for useful discussions. D. Giataganas
is supported by a Claude Leon postdoctoral fellowship and also participates in PEVE
2010 NTUA program. The work of Y. Lozano and M. Picos has been partially supported by the research grants MICINN-09-FPA2009-07122, MEC-DGI-CSD2007-00042 and COF10-03.
K. Siampos has been supported by the ITN programme PITN-GA-2009-237920, the ERC Advanced Grant 226371, the IFCPAR
CEFIPRA programme 4104-2 and the ANR programme blanc NT09-573739. Y. Lozano would also like to thank the Newton Institute and K. Siampos the University 
of Patras and the Theory Unit at CERN, for hospitality while part of this work was done. D. Giataganas would like to thank the Bethe Center for Theoretical Physics and
Physikalisches Institut, University of Bonn, and the Centre de Physique Théorique, Ecole Polytechnique for hospitality during the final stage of
this work.

\startappendix
%c
\Appendix{The $AdS_5\times Y^{p,q}$ background}

In this Appendix we collect some properties of $Y^{p,q}$ manifolds useful for the description of the baryon vertex in the $AdS_5 \times Y^{p,q}$ background. The Klebanov-Witten background is described thereof as a particular case\footnote{With the well-known subtleties regarding the periodicities.}. We also provide the detailed microscopical description of the baryon vertex in $AdS_5\times Y^{p,q}$ in terms of D3-branes expanding into a fuzzy 2-sphere.

\subsection{Some properties of the $AdS_5\times Y^{p,q}$ geometry}

In our conventions the $AdS_5\times Y^{p,q}$ metric reads
\begin{equation}
ds^2=R^2 \Bigl(ds^2_{AdS_5} + ds^2_{Y^{p,q}}\Bigr)=\frac{\rho^2}{R^2}dx^2_{1,3}+\frac{R^2}{\rho^2}d\rho^2+R^2 ds^2_{Y^{p,q}}~,
\end{equation}
with $R$ the radius of curvature in string units,
\begin{equation}
R^4=\frac{4\pi^4Ng_s}{{\rm Vol}(Y^{p,q})}~.
\end{equation}
%d1Next we collect some properties of the $Y^{p,q}$ geometries of \cite{gauntletts2s3} that will be relevant for our analysis. More details about these geometries can be found in the original paper.
For the $Y^{p,q}$ we use the canonical form of the metric \cite{se}, given by:
\begin{eqnarray}
\label{metric2}
   ds^2_{Y^{p,q}} &= & \frac{1-cy}{6}(\diff \theta^2+\sin^2\theta \diff \phi^2)
      + \frac{\diff y^2}{w(y)q(y)}
      + \frac{1}{36} w(y)q(y)(\diff \beta+c\cos\theta \diff \phi)^2 \nonumber \\
      && + \frac{1}{9}[ \diff \psi+\cos\theta \diff \phi
         +y(\diff \beta+c\cos\theta \diff \phi)]^2=\nonumber\\
   &=&(e^\theta)^2+(e^\phi)^2+(e^y)^2+(e^\beta)^2+(e^\psi)^2~,
\end{eqnarray}
where the f\"unfbeins read
\bea
&&e^{\theta}=\sqrt{\frac{1-cy}{6}}\;\diff\theta, \qquad\quad
e^{\phi}=\sqrt{\frac{1-cy}{6}}\;\sin\theta \diff\phi, \nonumber\\
&&e^y=\frac{1}{\sqrt{w(y)q(y)}}\diff y, \qquad
e^{\beta}= \frac{\sqrt{w(y)q(y)}}{6}(\diff\beta+ c\cos\theta \diff\phi), \nonumber\\
&&e^{\psi}=\frac{1}{3}\left( \diff\psi+\cos\theta \diff\phi+y(\diff\beta+c\cos\theta \diff\phi)\right),
\eea
with
\ben
w(y)  =  \frac{2(a-y^2)}{1-cy}~ ,\quad
q(y)  =  \frac{a-3y^2+2cy^3}{a-y^2}~ ,
\een
and the metric is normalized such that $R_{\alpha\beta}=4\, G_{\alpha\beta}$.
The ranges of the coordinates $(\theta,\phi,\psi)$ are
$0\leqslant \theta \leqslant \pi$, $0\leqslant \phi \leqslant 2\pi$ and $0\leqslant \psi \leqslant 2\pi$. The parameter $a$ is restricted to $0<a<1$.
By choosing this range the following conditions for $y$ are satisfied: $y^2 < a$, $w(y)>0$ and $q(y)\geqslant 0$.
The coordinate $y$ then ranges between the two smaller roots of the cubic equation $q(y)=0$, $y_1\leqslant y \leqslant y_2$.
For $c\ne 0$, $y$ can always be rescaled such that $c=1$ and the parameter $a$ can be written in terms of two coprime integers $p$ and $q$ as:
\be\label{apq}
a=\frac{1}{2}-\frac{p^2-3q^2}{4 p^3}\sqrt{4 p^2 -3 q^2}~.
\ee
In this case
\begin{eqnarray}
 y_1=\frac{1}{4p}\Bigl( 2p-3q-\sqrt{4p^2-3q^2}\Bigr)<0\, , \qquad
 y_2=\frac{1}{4p}\Bigl( 2p+3q-\sqrt{4p^2-3q^2}\Bigr)>0\, .
 \end{eqnarray}
Finally, $\beta$ ranges between $-2\pi (6l+c)\leqslant \beta \leqslant 0$, where
\begin{equation}
l=\frac{q}{3q^2-2p^2+p\sqrt{4p^2-3q^2}}~.
\end{equation}
Note that $\beta$ needs not be periodic in general.
The volume of the $Y^{p,q}$ can be written in terms of $p,q$ as
\ben
{\rm Vol}(Y^{p,q})=\frac{q(2 p +\sqrt{4 p^2-3 q^2})l \pi^3}{3 p^2}\, .
\een
The canonical metric (\ref{metric2}) takes the standard form
\be
\label{metrica1}
\diff s^2_{Y^{p,q}}=\diff s^2_{M_4}+(\frac{1}{3} \diff\psi+\s)^2~,
\ee
where the Killing vector $k^\mu=\delta^\mu_\psi$ is the Reeb vector and $ds^2_{M_4}$ is a local K\"ahler-Einstein metric with K\"ahler form
\be
\label{j4}
J=\half \diff\sigma= \frac{1-cy}{6}\sin\theta\, \diff \theta \wedge \diff \phi + \frac{1}{6} \diff y
\wedge (\diff \beta +c \cos\theta \diff \phi)\, , \\
\ee
satisfying
\begin{equation}
\int_{M_4} J\wedge J=\frac{3{\rm Vol}(Y^{p,q})}{\pi}\, .
\end{equation}
This local property of the metric will be useful in order to induce an instantonic magnetic flux proportional to the K\"ahler form.

Finally, the $AdS_5\times Y^{p,q}$ flux reads $F_5=(1+\star_{10}){\cal F}_5$, where
\begin{equation}
{\cal F}_5=4\,R^4\, d{\rm Vol} (Y^{p,q})
\end{equation}
and
\begin{equation}
d{\rm Vol}(Y^{p,q})=e^\theta\wedge e^\phi\wedge e^y\wedge e^\beta\wedge e^\psi=\frac{1}{108}(1-cy) \sin{\theta}d\theta\wedge d\phi\wedge dy\wedge d\beta\wedge d\psi
\end{equation}
$F_5$ is then such that
\begin{equation}\label{CS-flux}
\frac{1}{(2\pi)^4g_s}\int_{Y^{p,q}}F_5=N~.
\end{equation}

\subsection{The $AdS_5\times T^{1,1}$ case}

As shown in \cite{se} when $c=0$ the metric (\ref{metric2}) reduces to the local form of the standard homogeneous metric on $T^{1,1}$. Indeed, setting $c=0$ in (\ref{metric2}), rescaling to set $a=3$ and introducing the coordinates $\cos{\omega}=y$, $\nu=-\beta$ one gets
\begin{equation}
\label{T11}
ds_{T^{1,1}}^{2} = \frac{1}{9}\left[d\psi - \cos{\theta}d\phi - \cos{\omega}d\nu \right]^{2}
+ \frac{1}{6}\left(d\theta^{2} + \sin^{2}{\theta} d\phi^{2}\right)
+ \frac{1}{6}\left(d\omega^{2} + \sin^{2}{\omega}d\nu^{2}\right)~,
\end{equation}
which is the metric of the $T^{1,1}$ in adapted coordinates to its realization as a $U(1)$ bundle over $S^2\times S^2$ \cite{CDLO}, normalized such that $R_{\alpha\beta}=4\, G_{\alpha\beta}$. Note however that although it is possible to take the period of $\nu$ equal to $2\pi$ the period of $\psi$ is fixed to $2\pi$, so the manifold that is being described in the $c=0$ case is the $T^{1,1}/\mathbb{Z}_2$ orbifold.  Still, we can study the baryon vertex in $T^{1,1}$ as a particular case of $Y^{p,q}$ geometry if we account for the right periodicity of $\psi$ when relevant.

The K\"ahler form in the $T^{1,1}$ reads
\begin{equation}
J=\frac{1}{6}\Bigl(\sin\theta d\theta\wedge d\phi+\sin\omega\, d\omega\wedge d\nu\Bigr)
\end{equation}
and some properties used in the main text are
\begin{equation}
\label{quantJ}
\int_{S^2} J= \frac{2\pi}{3}\, , \qquad \int_{T^{1,1}}J\wedge J= \frac{3{\rm Vol}(T^{1,1})}{2\pi}\, ,
\end{equation}
where the volume of the $T^{1,1}$ is given by
\begin{equation}
{\rm Vol}(T^{1,1})=\frac{16\pi^3}{27}\, .
\end{equation}
Finally, the 5-form field strength is $F_{5} = (1+\star_{10}){\cal F}_5$, where
\begin{equation}
\label{F5T11}
\mathcal{F}_5 \equiv 4R^{4}~\textrm{dVol}\left(T^{1,1}\right)=\frac{R^4}{27} \sin{\theta}\sin{\omega}\, d\theta \wedge d\omega\wedge d\psi \wedge d\phi \wedge d\nu
\end{equation}
and satisfies:
\begin{equation}
\frac{1}{(2\pi)^4g_s}\int_{T^{1,1}}F_5=N~.
\end{equation}

\subsection{The baryon vertex in $AdS_5\times Y^{p,q}$ with a magnetic flux proportional to the K\"ahler form of the $S^2$}

The microscopical description of the baryon vertex in $AdS_5 \times Y^{p,q}$ in terms of D3-branes expanding into a fuzzy 2-sphere is complementary to a macroscopical D5-brane wrapped on the $Y^{p,q}$ with a magnetic flux proportional to the K\"ahler form on the $S^2$. This magnetic flux dissolves D3-brane charge, with the D3's spanned on the $(y,\beta,\psi)$ directions.

The DBI action for the D5-brane in the Sasaki--Einstein background \eqn{metric2} reads
\bea
S^{DBI}_{D5}=-T_5\int_{\mathbb{R}\times Y^{p,q}}d^6\xi\, e^{-\phi}\sqrt{-\det{P(G_{MN}+2\pi F_{MN})}}\ , 
\eea
where $M=(\mu;i)=(t,a;i)\ ,a=(y,\beta,\psi)\ ,i=(\th,\phi)$. Turning on a magnetic flux proportional to the 
K\"ahler form on the $S^2$ parameterized by $\theta$ and $\phi$ in  (\ref{metric2}) it is easy to prove that
\bea
&&\det{P(G_{MN}+2\pi F_{MN})}=G_{tt}IJ\ , \qquad {\rm with} \\
&&I=\det[(G+2\pi F)_{ij}]\ ,\qq J=\det[G_{ab}-G_{ai}(G+2\pi F)^{-1ij}G_{jb}]\ .\nonumber
\eea
The determinant $I$ can be easily computed, with the result
\bea
I=G_{\th\th}G_{\phi\phi}\left(1+(2\pi\cN)^2\varepsilon^2\right)\equiv G_{\th\th}G_{\phi\phi}\,\cI \ , \qq \varepsilon^2\equiv\frac{1}{(1-y)^2R^4}\ .
\eea
For the computation of $J$ we note that $G^{ab}G_{bi}=-\d_{a\psi}\d_{i\phi}\cos\th$.
The result reads
\bea
J=(\det[G_{ab}])^2\det\left[G^{ab}-\frac{\d_{a\psi}\d_{b\psi}\cos^2\th}{G_{\phi\phi}\left(1+(2\pi\cN)^2
\varepsilon^2\right)}\right]\equiv
\det[G_{ab}]\cJ\ .
\eea
Plugging these expressions in the DBI action we finally find
\bea
\label{DBI.macro.SE}
E_{D5}=N\frac{\rho_0}{16\pi}
\frac{\int_{y_1}^{y_2}dy(1-y)\int_0^\pi d\th\sin\th\sqrt{\cI\cJ}}{\int_{y_1}^{y_2}dy(1-y)}\ ,
\eea

\subsection{The microscopical construction}

In this appendix we show that the baryon vertex with magnetic flux that we have just discussed can be described at finite 't Hooft coupling in terms of D3-branes expanding into a fuzzy 2-sphere. The geometry of the fuzzy D5-brane is then given by the twisted product of the 3 dimensional manifold spanned by the $(y, \beta, \psi)$ directions and a fuzzy 2-sphere.

The DBI action describing the dynamics of $n$ coincident D3-branes spanned on the
$(y, \beta, \psi)$ directions and expanding onto the fuzzy $S^2$ parameterized by $\theta$ and $\phi$ in (\ref{metric2}) is given by ($I=1,2,3$)
\bea
\label{DBI.micro.SE}
S^{DBI}_{n D3}=-T_3\int d^4\xi \,{\rm Str}\left[e^{-\phi}\sqrt{-G_{tt}\tilde{\cI}\tilde{J}}\ \right]\ ,
\eea
where
\bea
\tilde{\cI}=\det Q^I{}_J\ ,\qq
\tilde{J}=\det P[G_{ab}+G_{a I}(Q^{-1}-\d)^{IJ}G_{Jb}]\ .
\eea
The determinant of $Q^I{}_J$ can be computed in a similar way as in the previous cases, and the result reads
\bea
&&Q^I{}_J=\d^I{}_J-\frac{\L_{(m)}}{2\pi}\ \varepsilon^{IK}{}_L\ X^L G_{KJ}\ ,\qq \L_{(m)}=\frac{2}{\sqrt{m(m+2)}}\ \Longrightarrow \\
%&&Q^i{}_j\simeq-\frac{\L_{(m)}}{2\pi}\varepsilon^{ik}\left(G_{kj}+\frac{2\pi}{\L_{(m)}}\varepsilon_{kj}\right)\Longrightarrow
&&\det Q^I{}_J\simeq\left(\frac{\L_{(m)}}{2\pi}\right)^2\varepsilon^{-2}\left(1+\left(\frac{2\pi}{\L_{(m)}}\right)^2\varepsilon^2\right)\ . \nonumber
\eea
Next, we consider the determinant $\tilde J$ and we note that
\bea
Q^I{}_J=\d^I{}_J-\frac{\L_{(m)}}{2\pi}\ \varepsilon^{IK}{}_L\ X^L G_{KJ}=
G^{IK}(G_{KJ}-\frac{\L_{(m)}}{2\pi}\varepsilon_{KJL}X^L)\Longrightarrow Q=G^{-1}\tilde{Q}\ .
\eea
Thus, we have to compute the inverse of $\tilde Q$, which in the macroscopical limit ($m\gg1$) reads
\bea
\label{InverseSU(2)}
&&\tilde Q_{IJ}=G_{IJ}+\varepsilon_{IJL}v^L\Longrightarrow
\tilde Q^{-1IJ}=\frac{1}{1+v^2}\left(G^{IJ}+v^Iv^J-\varepsilon^{IJ}{}_Kv^K\right)\ ,\\
&&\quad v^I=-\frac{\L_{(m)}}{2\pi} X^I\ ,\qq v^2=G_{IJ}\ v^Iv^J=\left(\frac{\L_{(m)}}{2\pi}\right)^2\frac{(1-y)^2R^4}{36}\ ,\nonumber
\eea
where the indices are raised using $G^{IJ}$. Next, we compute $(Q^{-1}-\d)^{-1IJ}$ which reads
\bea
(Q^{-1}-\d)^{-1IJ}\equiv(Q^{-1}-\d)^{-1I}{}_K\ G^{KJ}=\frac{1}{1+v^2}\left(-v^2G^{IJ}+v^Iv^J-\varepsilon^{IJ}{}_Kv^K\right)\ .
\eea
So, using the last equation and $G^{ab}G_{bi}=-\d_{a\psi}\d_{i\phi}\cos\th$ we find
\bea
\tilde J=(\det[G_{ab}])^2\det\left[G^{ab}-\frac{\d_{a\psi}\d_{b\psi}\cos^2\th}{G_{\phi\phi}\left(1+v^{-2}
\right)}\right]\equiv\det[G_{ab}]\tilde\cJ\ .
\eea
Putting all these ingredients together in \eqn{DBI.micro.SE} and
using  Eqn.(4.44) of \cite{Abe}; so to find the leading behavior for $m\gg1$, we find
\bea
\label{DBI.macro.SE1}
&&S^{DBI}_{n D3}=-T_3\int d^4\xi {\rm Str}\left[e^{-\phi}\sqrt{-G_{tt}\tilde{\cI}\tilde{J}}\ \right]
\simeq-T_3\frac{\L_{(m)}}{2\pi}\frac{m+1}{4\pi}
\int_{S^2}dS^2\int d^4\xi\left[e^{-\phi}\sqrt{-G_{tt}\tilde{\cI}\tilde{J}}\ \right] \nonumber\\
&&=-T_5\frac{m+1}{\sqrt{m(m+2)}}\int d^6\xi\left[e^{-\phi}\sqrt{-G_{tt}\tilde{\cI}\tilde{J}}\ \right]\Longrightarrow\nonumber\\
&&E^{DBI}_{nD3}=N\frac{m+1}{\sqrt{m(m+2)}}\frac{\rho_0}{16\pi}
\frac{\int_{y_1}^{y_2}dy(1-y)\int_0^\pi d\th\sin\th\sqrt{\tilde{\cI}\ \tilde{\cJ}}}{\int_{y_1}^{y_2}dy(1-y)}\ ,
\eea
which in the large $m$ limit reproduces the macroscopical result and like as the $T^{1,1}$ case it is given by
\eqn{DBI.macro.SE} with $m={\cal N}/3$, for which $(\cI,\cJ)\Leftrightarrow (\tilde{\cI}, \tilde{\cJ})$.

\subsubsection{The F-strings}

The CS action describing the dynamics of the $n$ coincident D3-branes takes the form
\bea
\label{CSmicroSE1}
S^{CS}_{nD3}=T_3\int{\rm Str} \Bigl(P[C_4]+i P[(i_X i_X)C_4]\wedge F\Bigr)=
-i\, T_3\int{\rm Str}\Bigl( P[(i_X i_X)F_5]\Bigr)\wedge A_t
\eea
$F_5$ reads, in Cartesian coordinates for the $S^2$
\bea
F^{(5)}_{y\beta\psi ij}=\frac{R^4}{27}\, (1-y) \varepsilon_{ijk} X^k
\eea
Substituting in the action we find that
\bea
\label{CSmicroSE2}
S^{CS}_{nD3}=-N\frac{m+1}{\sqrt{m(m+2)}}\int A\ ,
\eea
which exactly matches the macroscopical result  (\ref{tadpole}) in the large $m$ limit.

\bibliographystyle{cite}

\end{document}